\documentclass[
 superscriptaddress,
preprint,
 amsmath,amssymb,
 aps,
floatfix,
]{revtex4-2}

\usepackage{graphicx}
\usepackage{dcolumn}
\usepackage{bm}
\usepackage{hyperref}
\usepackage[mathlines]{lineno}

\begin{document}

\preprint{APS/123-QED}

\title{Inverse design and experimental realization of plasma metamaterials}

\author{Jesse A. Rodr\'iguez}
\email{jesse.rodriguez@oregonstate.edu}
\affiliation{Department of Mechanical Engineering, Stanford University, Stanford, CA 94305}
\author{Mark A. Cappelli}
\affiliation{Department of Mechanical Engineering, Stanford University, Stanford, CA 94305}

\date{\today}

\begin{abstract}
	We apply inverse design methods to produce two-dimensional triangular-lattice plasma metamaterial (PMM) devices which are then constructed and demonstrated experimentally. Finite difference frequency domain simulations are used along with forward-mode automatic differentiation to optimize the plasma densities of each of the plasma elements in the PMM to perform beam steering and demultiplexing under transverse magnetic polarization. The optimal device parameters are then used to assign plasma density values to elements that make up an experimental version of the device. Device performance is evaluated against both the simulated results and human-designed alternatives, showing the benefits and disadvantages of \textit{in-silico} inverse design and paving the way for future fully \textit{in-situ} optimization.
\end{abstract}

                    
\maketitle

\section{Introduction}
	
The inverse design technique is aptly named; it consists of the process of optimizing a device given an \textit{a-priori} set of performance metrics. In electromagnetically active systems, the technique becomes a type of constrained optimization problem with Maxwell's equations serving as the constraints. The performance metrics are encoded in an objective function that increases with favorable performance, \textit{e.g.} $L_2$ inner products of the electromagnetic (EM) field solutions for the current design iteration and the desired EM fields in regions of interest. This objective is a function of user-defined parameters that encode the domain/configuration of the device. This technique has been very fruitfully applied in EM systems over the last decade, leading to novel devices that perform functions so complex and at such high efficiencies that they would be impossible to realize through expert-level human design \cite{Su2020Spins, miller2013photonic, liu2013transformation, hughes2018adjoint, molesky2018inverse, christiansen2021inverse,  AndradeInverseGuide}. Some examples of such devices are high performance metalenses \cite{ChristiansenTunableLens,MeemInverseFlatLens}, metasurfaces \cite{ PestourieMetasurfacesDemultiplex, ChungTunableMetasurface}, photonic crystals \cite{ BorelTopologyWaveguide, minkov2020inverse, burger2004inverse, LinPhotonicDirac,  designsurvey}, and even optical computing devices that can perform functions such as boolean logic operations \cite{RodriguezPRA} and matrix-vector multiplication \cite{MatVecPenn}.
	
As mentioned above, in order to construct an objective that enables optimization of the device domain, the designer must come up with a parameterization scheme. Unfortunately, the configuration space of physically realizable devices is much smaller than the configuration space of devices that can be modeled, so the domain is parameterized in a manner that is in agreement with experimental facilities. Typically, given the limits of working with existing dielectric materials which have a fixed dielectric response, the domain is binarized and the parameters simply correspond to whether or not material with a given dielectric constant is present at a location in the domain \cite{piggott2015inversebinarized, su2018inversebinarized}. Of course, this limits the space of possible devices, but nevertheless this approach has yielded startlingly strong results. There is also some exciting recent work that involves the use of additive manufacturing to produce two-dimensional devices with a finite number of grayscale permittivity values within a certain interval through fine layering \cite{Grayscale}, greatly expanding the configuration space for inverse design devices. 
	
\begin{figure}[htbp]
	\centering
	\includegraphics[width=1\linewidth]{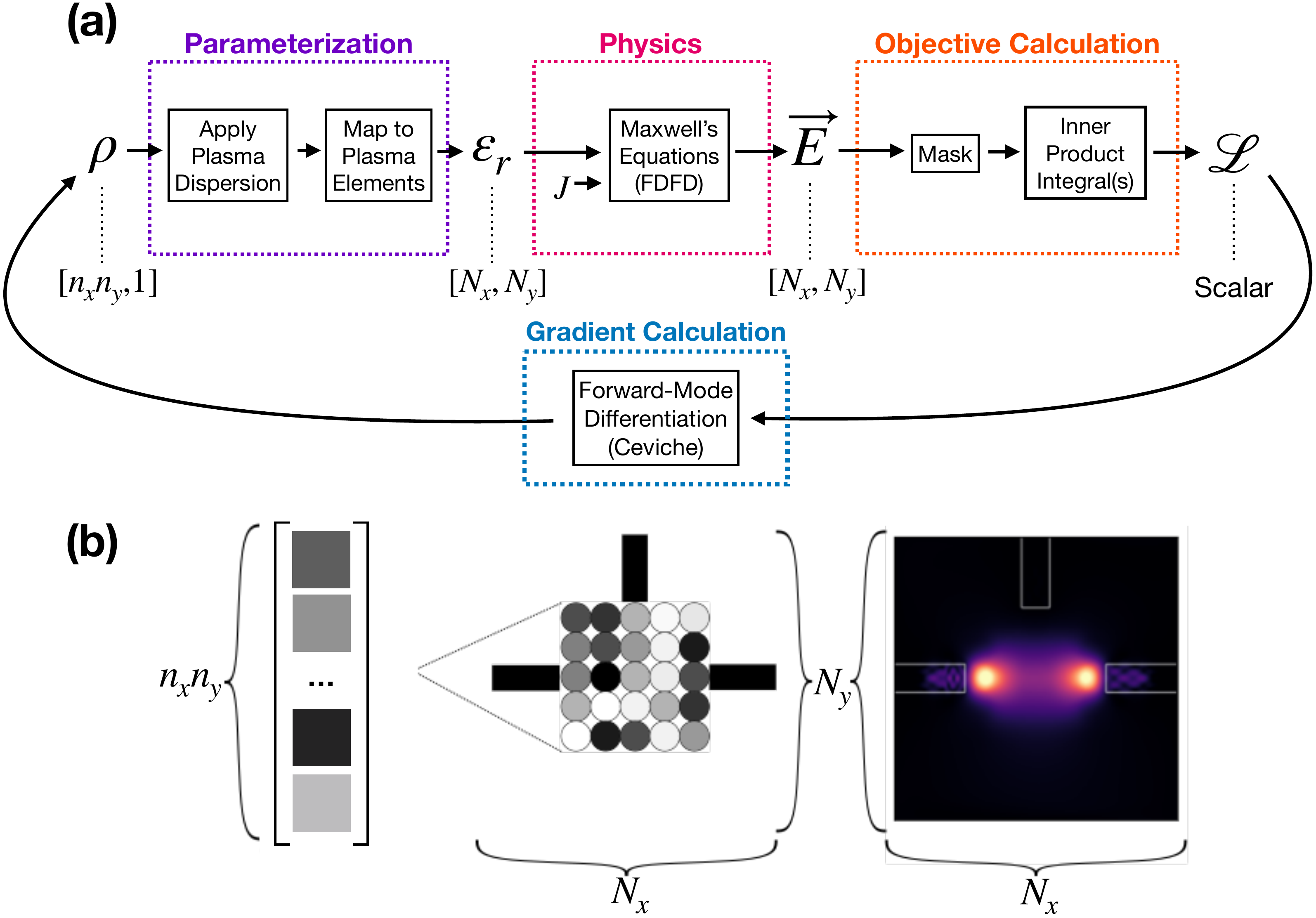}
	\caption{(a) Flow chart describing the algorithmic design of our PMM array. $J$ represents the modal source for the FDFD simulation. Examples of the training parameter vector $\rho$, permittivity matrix $\varepsilon_r$ and simulated $\textbf{E}$ fields for a simple PMM device are provided in (b).}
	\label{fig:invdes}
\end{figure}

In this and previous studies \cite{RodriguezPRA, RodriguezJPD}, we choose to encode our device domain as a plasma metamaterial (PMM), that is, as a domain filled with tunable plasma elements that can be reconfigured within a continuous range of permittivity values. A diagram that illustrates the inverse design algorithm in our context is shown in Figure \ref{fig:invdes}. In a PMM, each of the device parameters corresponds to the plasma frequency (proportional to the square root of the plasma density) of one of the elements in the device. The plasma density is varied in practice by tuning the power delivery to the plasma elements; in our case by increasing power supply voltage for a low-temperature gas discharge, but possible by many methods in practice such as tuning the power delivered to a resonator cavity or of the source of a laser-generated plasma. \textit{In-silico}, the domain is encoded as such and then a finite difference frequency domain simulation (FDFD) is computed using an autograd-compliant Maxwell solver called Ceviche \cite{ceviche}. 'Autograd-compliant' here means that by defining an objective that depends on the EM fields produced by the solver ($L_2$ inner products as mentioned before), numerical gradients can easily be computed with respect to the parameters that encode the domain using a widely-known python package. Ceviche supplies the gradient of the fields with respect to the input parameters via forward-mode differentiation, a technique where the Jacobian is computed during the forward pass, e.g. as the solver progresses to produce the field solutions. The resulting gradient is then used to iteratively optimize the device. 
	
\begin{figure}[htbp]
	\centering
	\includegraphics[width=1\linewidth]{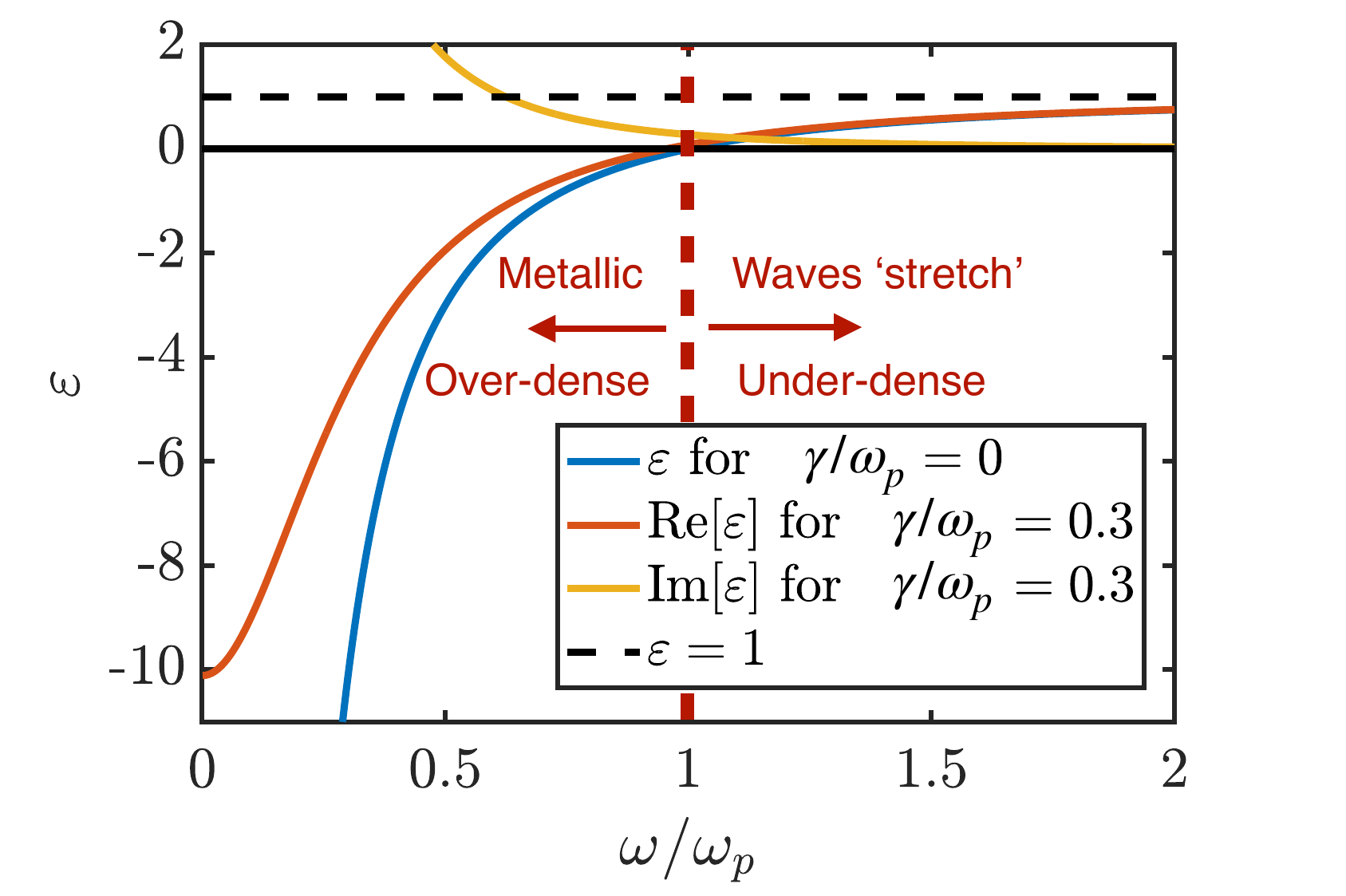}
	\caption{Permittivity of non-magnetized plasma vs incident wave frequency normalized by the plasma frequency.}
	\label{fig:plasmadisp}
\end{figure}

The choice to encode the device as a PMM has a number of advantages that are a direct result of the unique dielectric response of the plasma. Figure \ref{fig:plasmadisp} shows the way the dielectric permittivity of a non-magnetized plasma varies with incident wave frequency. The permittivity (neglecting the ion's response) is of the following form, known as the Drude permittivity:
\begin{align*}
	\varepsilon=1-\frac{\omega_p^2}{\omega^2+i\omega\gamma}
\end{align*}
where $\gamma$ is the collision/damping rate, $\omega_p^2=\frac{n_ee^2}{\varepsilon_0m_e}$ is the plasma frequency squared, $n_e$ is the electron density, $e$ is the electron charge, $m_e$ is the electron mass, and $\varepsilon_0$ is the free-space permittivity. We can see from this functional form that in the collisionless limit, the plasma permittivity can be tuned continuously in the interval $(-\infty,1)$ by varying the plasma density for a given operating frequency $\omega$. By choosing an operating frequency that is close to the plasma frequency of the elements, we can easily make drastic changes to their electromagnetic response; \textit{i.e.} that of a metallic element or a near-zero-index medium. This dispersion relation not only makes the configuration space infinite as opposed to a binarized parameterization scheme, but much more importantly, this allows a single array of plasma elements to perform several functions, and since the plasma elements can be quickly deactivated, you also can switch the device on and off very quickly $\sim 10$ ms or less. More detail about switching times is given in the methods section below.
	
Most importantly, since all the parameters can be varied precisely in the physical device, it is possible to perform the inverse design algorithm entirely \textit{in-situ}. This expedites the process of designing these devices by eliminating the need for expensive numerical simulations and also removes the errors associated with inaccurate modeling of the device elements. Fully \textit{in-situ} inverse design is just now becoming experimentally feasible \cite{InsituBackprop}, and the use of plasma in inverse design devices offers high-precision, continuous reconfigurability of device elements, attributes which are ideally suited for \textit{in-situ} optimization. In this study, we will examine the merits and disadvantages of creating inverse design devices the conventional way, by optimizing them \textit{in-silico} and then seeing how they perform experimentally. The development of the experimental platform which is detailed in this study will enable the transition to \textit{in-situ} inverse design in short order, but first we must examine how our best \textit{in-silico} optimization efforts translate to experimental results. In the following sections we will numerically optimize and experimentally demonstrate a two-dimensional PMM that can perform both beam steering and demultiplexing.

\section{Methods}
\subsection{Computational Inverse Design}
\begin{figure}[htbp]
		\centering
		\includegraphics[width=1\linewidth]{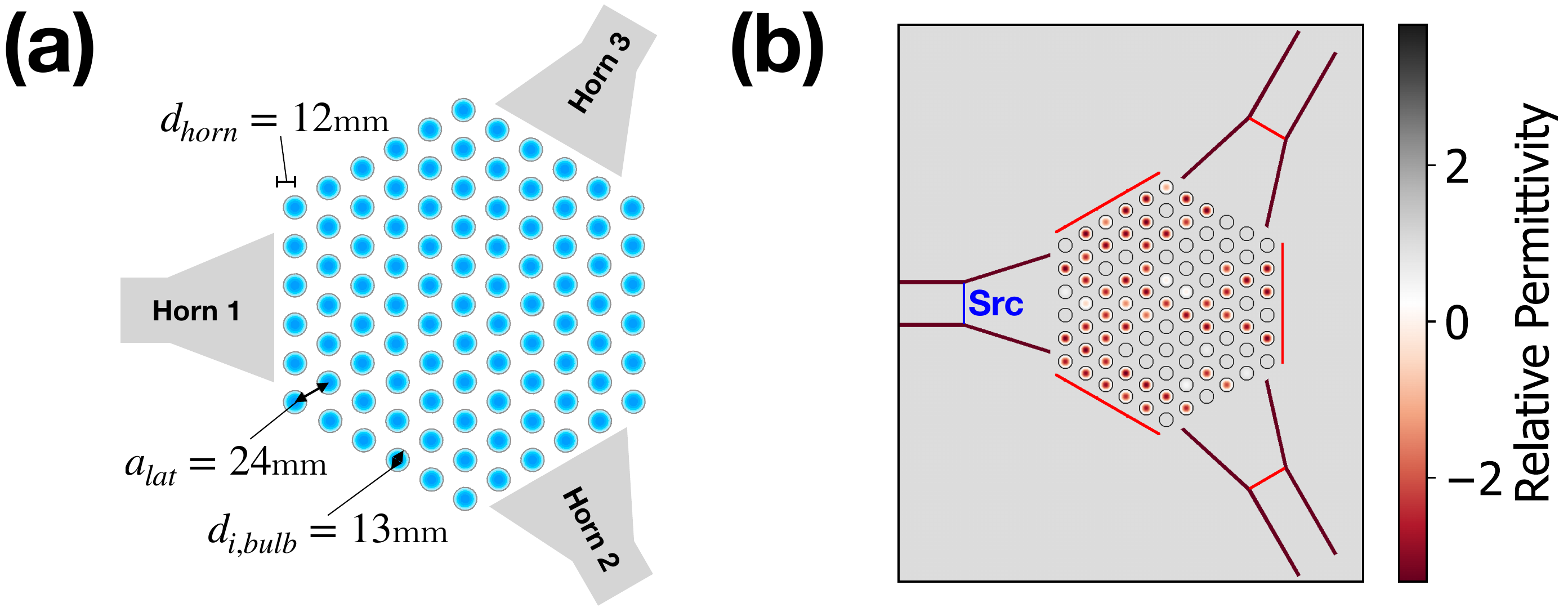}
		\caption{(a) Schematic of PMM and (b) relative permittivity of the simulation domain with source and objective evaluation locations denoted in blue and red respectively. The domain is shown with an arbitrary device configuration as an example.}
		\label{fig:schem}
	\end{figure}

Figure \ref{fig:schem}(a) provides a diagram of our PMM device. The device has a triangular lattice with lattice spacing $a=24$mm and is in the shape of a hexagon with six elements on each side, yielding 91 elements total. The outer diameter of the $\varepsilon=3.8$ quartz tubes is 15mm and the inner diameter is 13mm. In figure \ref{fig:schem}(b), the simulation domain is shown with an arbitrary average plasma density distribution among the elements, where each domain parameter corresponds to the space-averaged plasma frequency of one of the elements. The elements are modeled as either spatially uniform or as having a density profile that is axisymmetric and a zeroth-order Bessel function of the first kind of radius, with the first root at $r=6.5$mm. In the collisional model (which is also non-uniform), the simulated elements have a collision frequency of $\gamma=1$ GHz. The blue line indicates where the source is located in the domain, and the red lines indicate the locations in the domain where the objective function is evaluated; we refer to these as 'probe slices'. The probe slices surrounding the crystal are present to discourage leaking out of the device, while the probe slices within the exit horns either promote propagation into the correct horn or discourage propagation into the incorrect horn. The relative permittivity of the input and output horn walls was set to $\varepsilon=-1000$ to serve as a lossless metal, and the horn geometry was chosen to match our experimental microwave horns (A INFO LB-20180 2-18 GHz).
	
The basic idea is that a modal source is introduced at the input horn and allowed to scatter through the training region before collecting at the desired output horn. In the computational optimization phase, the fields are propagated through the domain via Ceviche. As mentioned earlier, Ceviche solves Maxwell's equations using a FDFD solver, and therefore all computed devices described here represent the steady state solution achieved after some characteristic time. The simulation domain was discretized using a resolution of 50 pixels per lattice constant $a$ and in each case presented dimensions of $N_x=1000$ and $N_y=1200$ pixels. The final device configurations were tested at the higher resolution of 150 pixels$/a$ and yielded results that were qualitatively identical. A perfectly-matched boundary layer (PML) $2a$ in width was applied along the domain boundaries. The simulated $\mathbf{E}$ fields are masked to compute integrals along the probe slices within the problem geometry. These integrals are used to calculate the objective $\mathcal{L}(\rho)$. Our simulation tool, Ceviche, is then used to compute numerical gradients of the objective with respect to the training parameters via forward-mode differentiation \cite{ceviche}. We then use the Adam optimization algorithm \cite{kingma2014adam} (gradient ascent with momentum, in essence) to iteratively adjust $\rho$ and thereby maximize $\mathcal{L}$. Optimization was conducted with learning rates ranging from $0.01-0.05$, and the default Adam hyperparameters $\beta_1=0.9$, and $\beta_2=0.999$.
	
The polarization of the input source has a strong effect in devices of this nature, either \textbf{$E_z$} ($E$ out of the page), which we call the transverse magnetic (TM) polarization, or \textbf{$E_x$} ($H$ out of the page), which we call the transverse electric (TE) polarization, with the latter case benefiting (and suffering, depending on the desired device functionality) from the presence of localized surface plasmon (LSP) modes \cite{righetti2018enhanced, tunableppc, SakaiNPerm, SakaiChain}, while the former makes more direct use of dispersive and refractive effects \cite{tunableppc, guideppc, ppc}. In our previous studies \cite{RodriguezPRA, RodriguezJPD}, we see that for devices which seek to preserve an input signal like those we present here, the TE polarization leads to poor performance, so we choose to focus on the TM mode alone in this study. Thus, we will not expect to see any small, sub-wavelength-scale field structures around our plasma elements or the large transmission losses which are indicative of the presence of LSP modes. 
	
To summarize, the PMM optimization problem can be expressed as
	\begin{align*}
		&\max_{\rho} \quad  \mathcal{L}(\mathbf{E}) \\
		&\text{given} \quad \nabla\times\frac{1}{\mu_0}\nabla\times \mathbf{E}-\omega^2\varepsilon(\omega,\rho)\mathbf{E}=-i\omega J.
	\end{align*}
where $\mathcal{L}$ is the objective composed of a set of $L_2$ inner product integrals of the simulated field with the desired propagation mode, $\textbf{E}$ is the electric field, $\rho$ is an $n$-dimensional vector (where $n$ is the number of elements) that contains the parameters that set the plasma frequency of each of the PMM elements, $\mu_0$ is the vacuum permeability, $\omega$ is the field frequency, $\varepsilon(\omega,\rho)$ is the spatially-dependent permittivity that is encoded by $\rho$, and $J$ is a current density used to define a fundamental modal source at the input horn. In practice, the permittivity distribution among the plasma elements is controlled by varying the discharge current (which therein alters the plasma density) in each of the PMM elements according to the Drude model (discussed above in the introduction), which is where the dependence on $\omega$ arises.
	
The beam steering objective is the $L_2$ inner product of the simulated field with an $m=1$ propagation mode (in order to preserve the input mode) evaluated at the probe in the desired exit minus the sum of the integrated field intensity at the incorrect exit and the other probe slices to discourage loss:
	\begin{equation*}
		\mathcal{L}_{steer}=\int E\cdot E_{m=1}^*dl_{\text{desired exit}}-\sum_{\text{prb}\neq\text{desired exit}}\int |E|^2dl_{\text{prb}}.
	\end{equation*}
The optimization objective for the demultiplexer is quite different from the beam steering case since the objective must take into account two separate simulations as the permittivity of the elements is frequency-dependent. With this in mind, the demultiplexer objective is:
	\begin{align*}
		\mathcal{L}_{mp} = &\left(\int E_{\omega_1}\cdot E_{m=1}^*dl_{\omega_1\text{ exit}}\right)\left(\int E_{\omega_2}\cdot E_{m=1}^*dl_{\omega_2\text{ exit}}\right)\\
        &-\int
		|E_{\omega_1}|^2dl_{\omega_2\text{ exit}}-\int |E_{\omega_2}|^2dl_{\omega_1\text{ exit}}\\
		&-\sum_{\text{prb}\neq\text{wvg exits}}\left[\int |E_{\omega_1}|^2dl_{\text{prb}}+\int |E_{\omega_2}|^2dl_{\text{prb}}\right],
	\end{align*}
where $E_{\omega_1}$ is the simulated field for the first frequency and $E_{\omega_2}$ is the simulated field for the second frequency. The first term in the objective is a product so that it ensures that the fields reach the exit horn for both operating frequencies (therein avoiding local minima where one gets good performance for just one of the two frequencies), the second two terms discourage leakage into the incorrect exits for each frequency, and the final term discourages leakage out of the device for both operating frequencies. For both objectives, each term is normalized by its value in the initial design iteration.
	
The maximum plasma frequency of the elements is set to be 8 GHz in the computational optimization procedure as that was our conservative estimate for the nominal maximum plasma density of our experimental sources. The optimal parameters from the computational inverse design process are then used to set the configuration of the experimental device according to the mapping defined below in Section \ref{sec:Exp}
	
\subsection{Experimental Realization}\label{sec:Exp}
The experimental plasma elements in this study are custom-manufactured UV germicidal discharge tubes made of high-purity quartz ($\varepsilon=3.8$) with an inner tube diameter of 13mm and an outer diameter of 15mm. The tubes are filled with argon gas at a fill pressure between 200 and 300 Pa, with a small amount of Mercury present in each discharge to provide the UV radiation. The use of UV discharge tubes is necessary because other types of off-the-shelf discharge tubes (for lighting applications) often have a fluorescent coating applied to the glass that is opaque to microwave-range radiation. When operating, the tubes are estimated to have a gas temperature between 315 and 330 K according to infrared thermometer measurements. The discharges are designed to operate with alternating current (AC) at a frequency between 20 and 50 kHz depending on the ballast circuit that is used. The nominal operating current and voltage is 24V and 0.6A, for a power of $\approx$ 15W. To push the discharges to higher plasma densities, we drive them with 24W DC ballasts (Beasun RL15-425-18D24), each having its own dedicated programmable DC power supply (Longwei LW-3010). The DC power supplies are connected to a RS485 serial communication bus (with maximum 32 power supplies per RS485 bus) and accept MODBUS protocol serial commands from a custom python library via a USB to RS485 converter. A schematic and photograph of the experimental scheme is included below in Figure \ref{fig:exp}. In practice, the discharges can be tuned through a fairly large range of plasma frequencies by setting the current of the DC power supply to its maximum value and then limiting the output voltage. The discharges typically operate from a maximum DC power supply voltage of 32 V down to about 6 V and then fail to maintain the discharge at a plasma frequency of about 2.4-4 GHz. We refer to this as the 'voltage tuning' mode. If the power supply is set to maximum voltage and the current is instead limited, a different operating mode can be reached, allowing us to access lower plasma frequency values all the way down to 0.4 GHz before the discharge extinguishes. For the high power ballasts used in this study, we achieve steady operation in the current tuning mode from approximately 60 mA to 200 mA.
	
	\begin{figure}[htbp]
		\centering
		\includegraphics[width=1\linewidth]{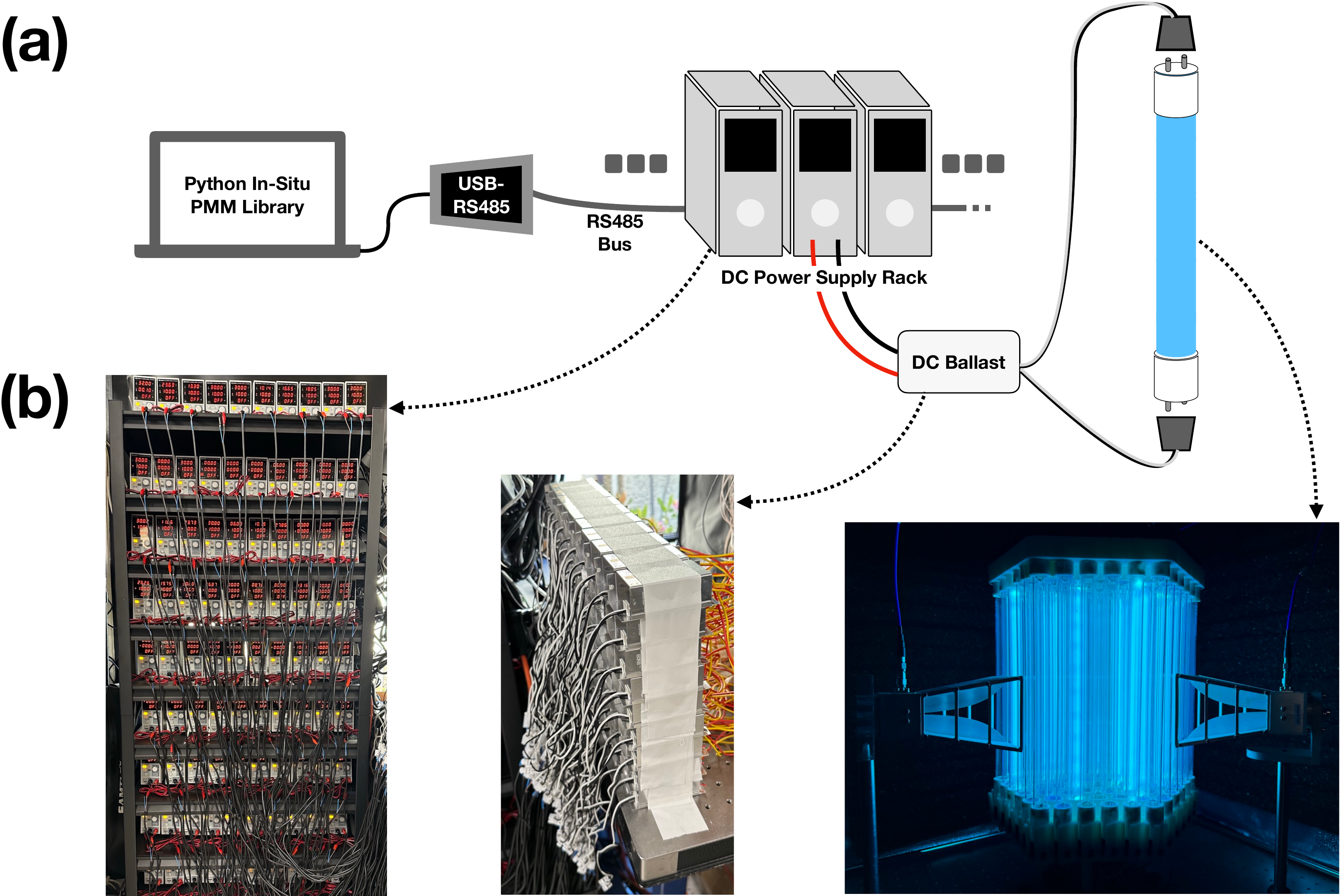}
		\caption{(a) Schematic of experimental PMM system and (b) photographs of the different components of the apparatus including the DC power supplies (left), the ballasts (middle), and the PMM itself activated in an arbitrary configuration (right).}
		\label{fig:exp}
	\end{figure}
	
In order to use the parameters from the \textit{in-silico} optimization which correspond to the plasma frequency of each of the plasma elements, we need to generate a mapping between the plasma frequency and the power supply settings that produce that frequency. This is quite challenging as we do not know the precise manufacturer fill pressure or gas temperature and the plasma density in our operating range is too small to use optical diagnostics for direct measurements. To combat this and obtain an approximate mapping from the power supply settings to plasma frequency, we use a 0-D kinetic simulation of the electron energy distribution function (EEDF) within a plasma discharge, BOLSIG+ \cite{BOLSIG, SigloDatabase}, assuming a known reduced electric field, $E/n$, where $n$ is the estimated neutral gas number density. Our estimate for the neutral gas number density is obtained using the ideal gas law along with estimates of the fill pressure and neutral gas temperature. We measure the time-varying voltage across and current through the plasma discharge for a range of operating conditions. The RMS voltage, estimated neutral gas number density, and estimates of the length of the discharge enables estimates of $E/n$. The EEDF determined using BOLSIG+ yields the electron mobility. The product of the mobility and electric field provides the space-averaged electron drift velocity $u_d=\mu_e E$ within the discharge. Finally, we can estimate the electron density using the measured current; $n_e = I_{RMS}/eu_dA_{tube,i}$, yielding the plasma frequency. Here, $A_{tube,i}$ is the internal cross sectional area of the discharge tube. We note that this is complex path to determine the plasma frequency, where several sources of uncertainty may be encountered.
	
Although the discharge frequency is much higher than the plasma recombination rate and as such the plasma density is quasi-steady, we have confirmed via high-speed photography that at the peak of the voltage cycle the plasma has the structure of a DC glow discharge with cathode and Faraday dark spaces clearly visible. At low-plasma density operating conditions, we also observe striations in the positive column of the discharge. This means that when considering the effective electric field applied to the plasma that we feed to the BOLSIG+ kinetic model, we have to take into account the cathode fall voltage and be sure that we use the correct discharge length. For our discharges, the discharge length varies between 190mm in the current-limited mode and 230mm in the voltage limited mode, and we estimate the cathode fall voltage to be somewhere between 6 and 11 V \cite{CathodeFall}. Taking all of these considerations into account, we perform a sweep of all parameters (fill pressure, temperature, cathode fall voltage) and perform a one-term polynomial fit that allows us to sweep between the highest and lowest-density cases by varying a parameter $k\in[0,1]$. Since the plasma frequency values appeared to be slightly too large for these types of discharges (owing to the many uncertainties in determining plasma density), we also added a scale factor $S$ that would allow us to adjust the fit to lower values over the entire tuning range. The quasi-experimental fit is included below in Figure \ref{fig:bulbtune}.
	
	\begin{figure}[htbp]
		\centering
		\includegraphics[width=1\linewidth]{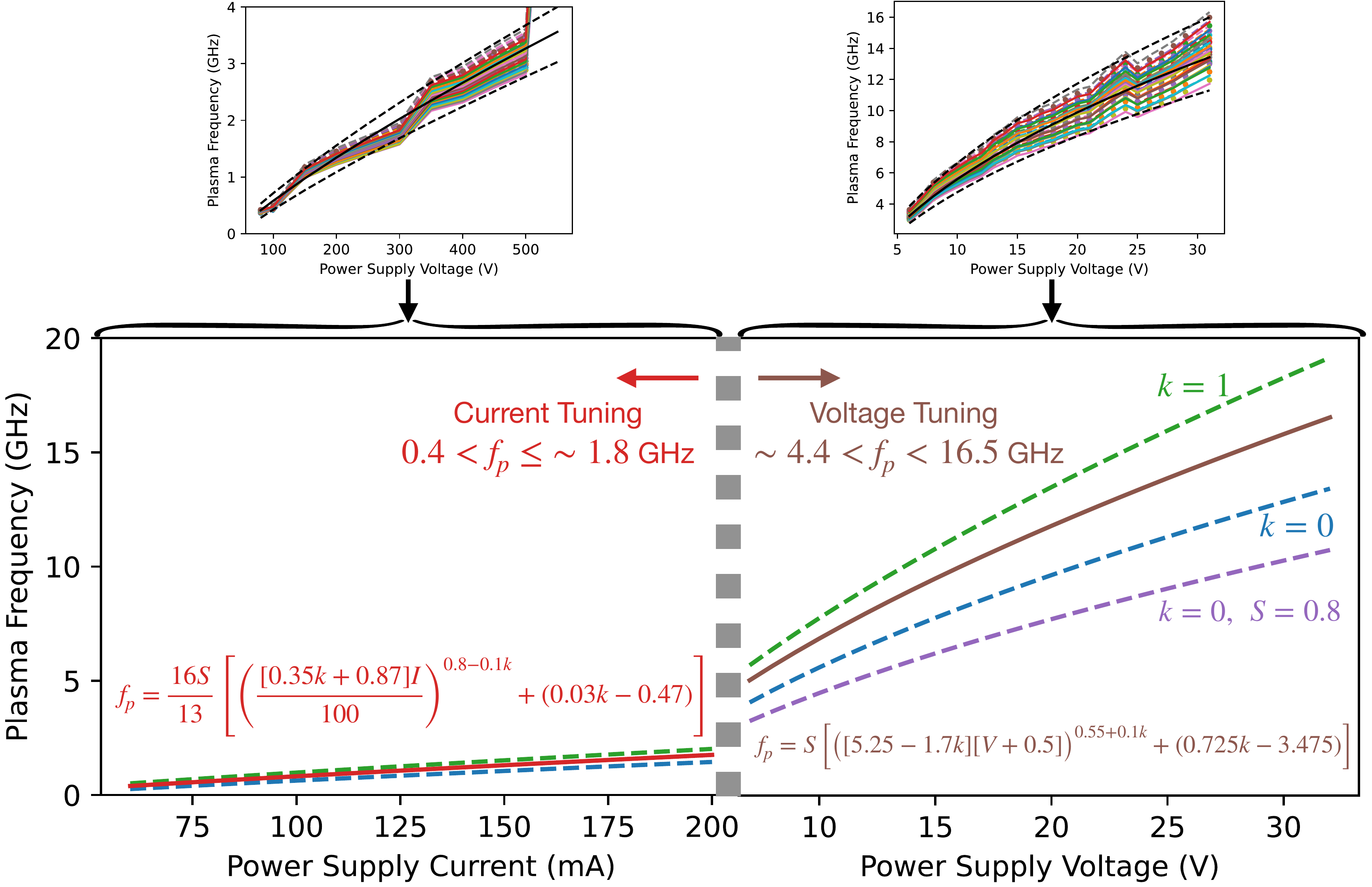}
		\caption{The quasi-experimental mapping for plasma tuning (bottom) and the plots including all the cases from the parameter sweep in BOLSIG+ that form the basis for the mapping (top).}
		\label{fig:bulbtune}
	\end{figure}

In addition to the high-speed photography images mentioned above, we have also conducted Schottky diode measurements of the transient dielectric response of the discharges as they are activated and deactivated, showing that the rise time of the dielectric response is $\sim10$ ms and the fall time is $\sim60$ ms. Due to the current configuration of the device where each RS485 bus hosts up to 25 power supplies, the actual time to set and activate the array is much longer, about 7 seconds, since each discharge is set and activated in series to avoid cross-talk within each RS485 bus. In practice, should a user need faster actuation, they need only parallelize the computational array setting procedure and use more RS485 buses. The theoretical limit is to have one cpu process communicating directly with a single power supply at a time with as many processes running in parallel as necessary. Since the power supplies operate at 9600 baud, each message is 8 bytes, and each actuation requires two messages (set voltage/current and activate), the total time added by the communication protocol is about 15 ms, so the time to switch the entire array on could be lowered to $\sim25$ ms and the deactivation time could be reduced to $\sim75$ ms. Temporal modulation of the plasma frequency of an already-running discharge could occur at a maximum of about 75 Hz. Of course, more sophisticated communication protocols and interconnects along with specially designed power supplies and ballast circuits could push these actuation times down further. Experimental data supporting these claims about the discharge structure and switching time can be found in the appendix of reference \cite{RodriguezThesis}.
 
To perform the measurements described in the Results section, we first construct the apparatus as pictured in Fig. \ref{fig:exp}, place the array into a custom-made anechoic chamber consisting of a box with microwave absorber panels attached to the interior surfaces, and then carry out an initial warm up of the device by turning every element in the device on for 15 seconds of every minute over a period of 10 minutes. This warm-up procedure makes sure that the temperature of the discharge tube will be steady throughout the measurements and it also allows the 3D-printed scaffolding to reach its steady operating temperature such that it won't flex/deform between measurements. Once the warm-up period is concluded, the PMM is activated according to the optimal parameters from the \textit{in-silico} optimization procedure for 15 seconds of each minute for as many minutes as it takes for all measurements to be collected. As stated before, the PMM device can activate and set the operating conditions for 91 elements in about 7 seconds. Python scripts are constructed to ensure that the PMM is not allowed to cool to room temperature by failing to maintain the 15 sec/min duty cycle.
	
For each measurement, $S_{21}$ and $S_{31}$ are measured by a Rohde and Schwarz ZNB40 4-port vector network analyzer that was calibrated using a Rohde and Schwarz ZN-Z54 calibration unit. The $S$-parameters are collected in 10,000 points from 2-12 GHz with an averaging factor of 10. For each device objective, we test the optimal parameters for three different plasma element models (uniform density profile, Bessel function profile, and Bessel function profile with collisions, see ref. \cite{RodriguezJPD} for more details regarding the plasma model) and sweep the fitting parameters $k$ and $S$ to attempt to find the correct plasma frequency mapping. This corresponds to about 30 measurements for each objective, and the best / most informative case is displayed below in Section \ref{sec:results}.

\section{Results}\label{sec:results}
\subsection{Beam Steering}
The simulated and experimental results for the beam steering objective where a source delivers power to one of two receiving ports are presented below in Figure \ref{fig:Wvg}. During the \textit{in-silico} optimization, the plasma density within each discharge is modeled as being either uniform with no collisions, non-uniform with no collisions, or non-uniform with collisions with $k$ and $S$ used to parameterize the relationship between the plasma density and operating voltage or current (see Figure \ref{fig:bulbtune}).
	
	\begin{figure}[htbp]
		\centering
		\includegraphics[width=1\linewidth]{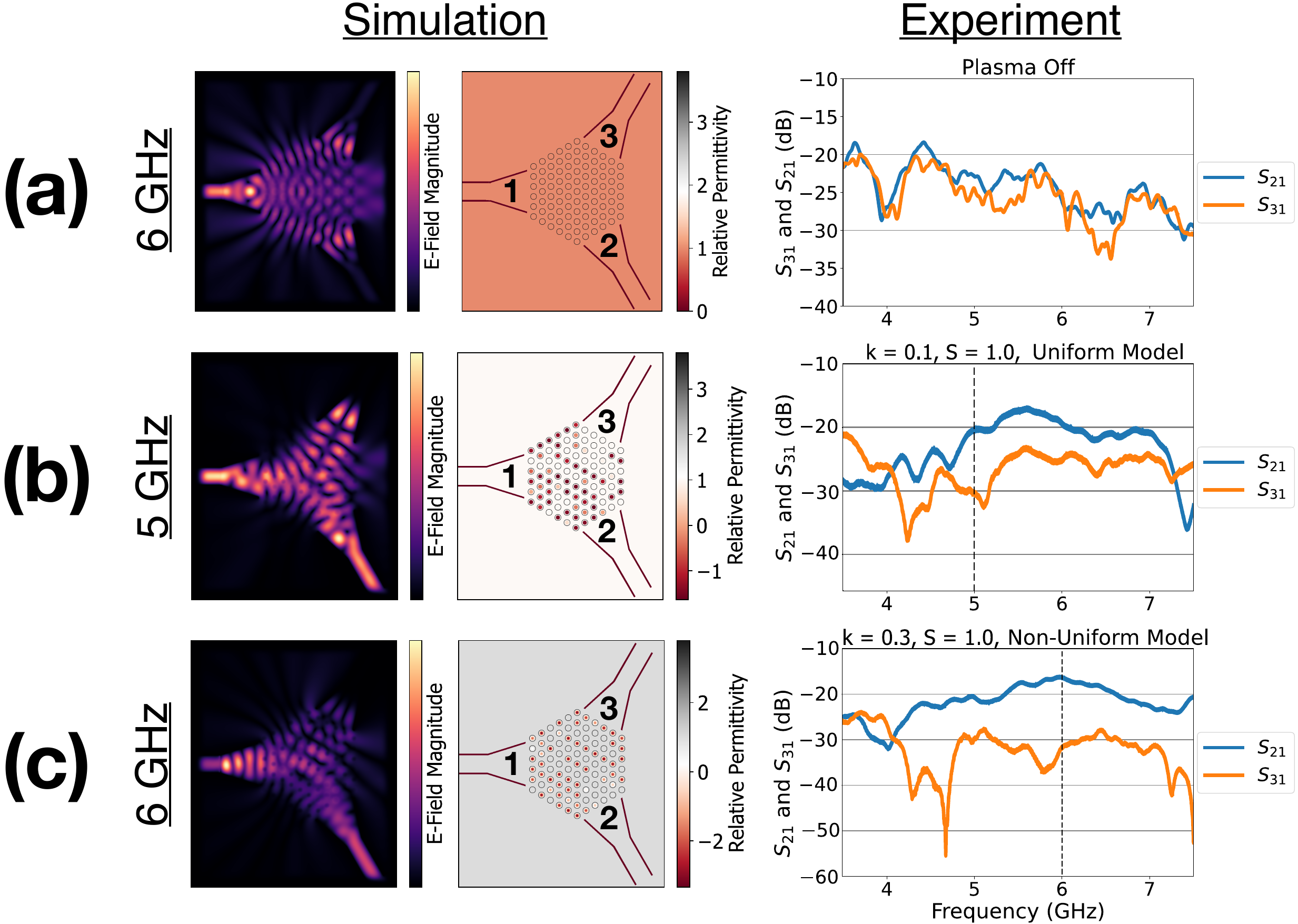}
		\caption{Field simulations $|\textbf{E}_z|$ and optimal device domains $\Re[\varepsilon]$ (left) along with experimental transmission spectra (right) for (a) the PMM device with all the elements inactive, (b) the 5 GHz operating frequency optimal beam steering design and, (c) the 6 GHz operating frequency optimal beam steering design. The vertical dashed black lines indicate the design frequency in (b) and (c), and the fit parameters $k$ and $S$ are given along with the plasma element model used (see Methods for details).}
		\label{fig:Wvg}
	\end{figure}
	
As expected, when all elements are inactive (Fig.\ref{fig:Wvg}(a)) the spectra of both ports are nearly identical within about 5 dB of one another at all frequencies. The small differences between the two measured spectra (right frame in Fig.\ref{fig:Wvg}(a)) are due to minor defects in the device scaffold and in the alignment of the microwave horns. In both the 5 and 6 GHz active configurations, the transmission in port 2 (the desired port) is nearly maximal at the frequency for which the device was optimized and we have about 10-15 dB isolation between the ports. In the 5 GHz configuration in particular, there is a drop in transmission in port 3 at almost precisely the operating frequency predicted by the simulations assuming a uniform plasma model with a discharge voltage mapping to plasma density defined by $k=0.1$ and $S=1.0$. In the 6 GHz plasma active case we see even better isolation at another frequency point, with over 30 dB isolation at about 4.7 GHz with a transmission in port 2 almost as high as at the operating frequency. We discuss the possible cause(s) for this and its significance in Section \ref{sec:disc}. In both cases, the power supply setting to plasma density fitting parameters that led to the best performance ($k\approx0.2$ and $S=1.0$), suggest that the actual maximum plasma density of the discharges when driven by their power supplies is is about 14 GHz (see Figure \ref{fig:bulbtune} for details) although prolonged operation at this power level causes damage to the discharge.
	
\subsection{Demultiplexer}
The simulated and experimental results for the demultiplexer objective are presented below in Figure \ref{fig:Demult}. Here, we use the same device as that above, but seek to separate different source frequencies. For example, in panel (b) of Figure \ref{fig:Demult}, we direct 5.5 GHz and 6 GHz content sourced at port 1, to ports 2 and 3 respectively.  
	
	\begin{figure}[htbp]
		\centering
		\includegraphics[width=1\linewidth]{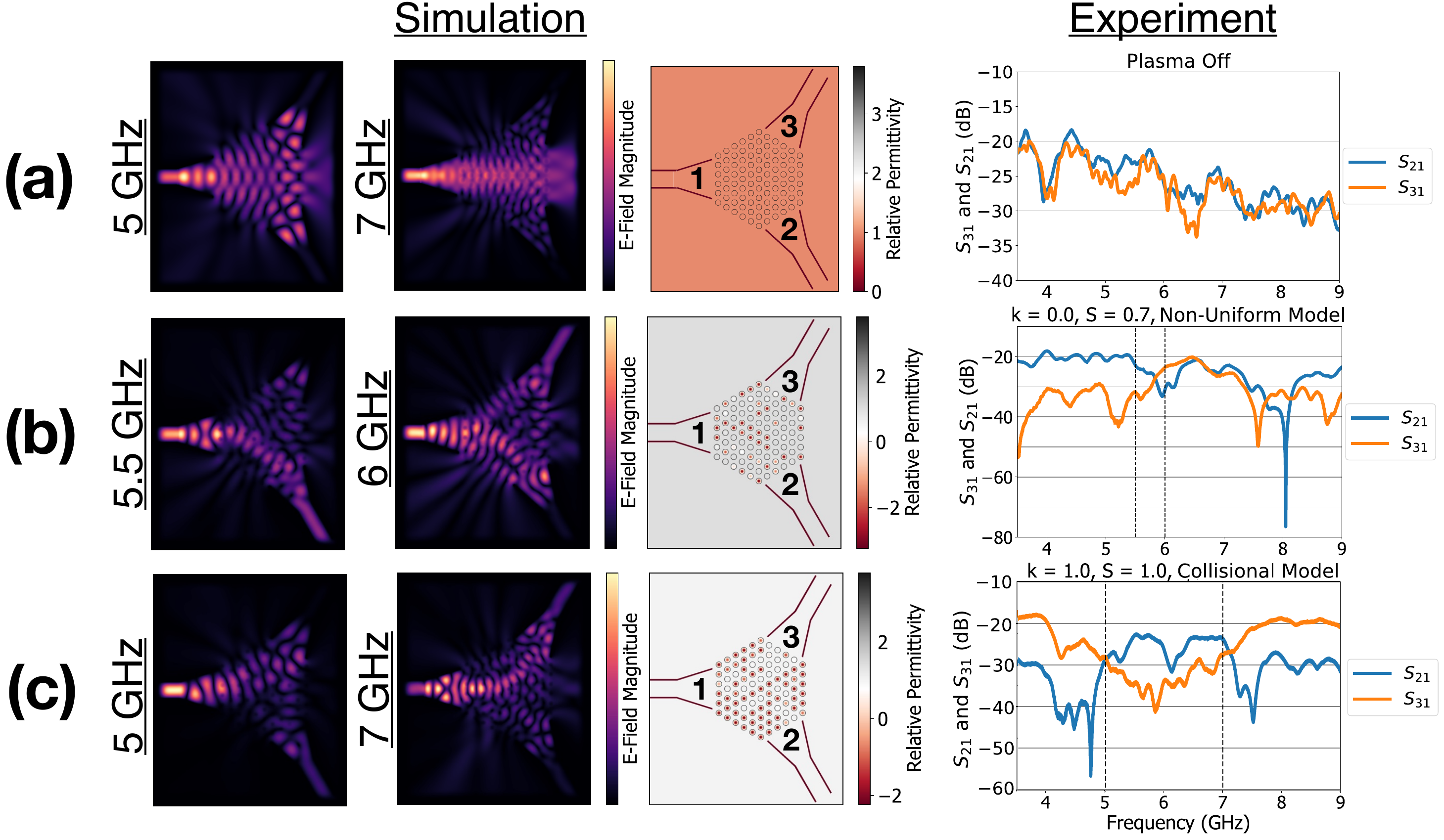}
		\caption{Field simulations $|\textbf{E}_z|$ and optimal device domains $\Re[\varepsilon]$ (left) along with experimental transmission spectra (right) for (a) the PMM device with all the elements inactive at 5 and 7 GHz, (b) the 5.5 / 6 GHz optimal demultiplexer design and, (c) the 5 / 7 GHz optimal demultiplexer design. The vertical dashed black lines indicate the design frequencies in (b) and (c), and the fit parameters $k$ and $S$ are given along with the plasma element model used (see Methods for details).}
		\label{fig:Demult}
	\end{figure}

Figure \ref{fig:Demult}(a) shows the baseline inactive plasma case, with experiments once again indicating that ports 2 and 3 receive similar signals (within 3-5 dB) over a wide range of frequencies. With active plasma cases, we find the best performance for the non-uniform model using $k=0.0$ and $S=0.7$ (similar to the beam steering objective) for the 5.5 / 6 GHz demultiplexer (Figure \ref{fig:Demult}(b)), but for the 5 / 7 GHz demultiplexer (Figure \ref{fig:Demult}(c)) we see the transmission characteristics we want for the collisional plasma element model at $k=1.0$, $S=1.0$. In the 5.5 / 6 GHz case, the transmission spectrum appears as one would expect in the region of the operating frequencies, with the transmission of the two ports crossing at the midpoint between the operating frequencies, where we see about 10 dB isolation in the correct direction. Interestingly, the same configuration performs the demultiplexer objective at the frequencies of $\sim$5.25 GHz and 8.05 GHz with up to 40 dB isolation, hinting at the potential of the device if the optimization was to be carried out \emph{in-situ}. The 5 / 7 GHz demultiplexer does not perform its objective properly at the design frequencies, but instead, performs very well at the slightly shifted frequencies of about 5.6 and 7.4 GHz. 
	
	\begin{figure}[htbp]
		\centering
		\includegraphics[width=1\linewidth]{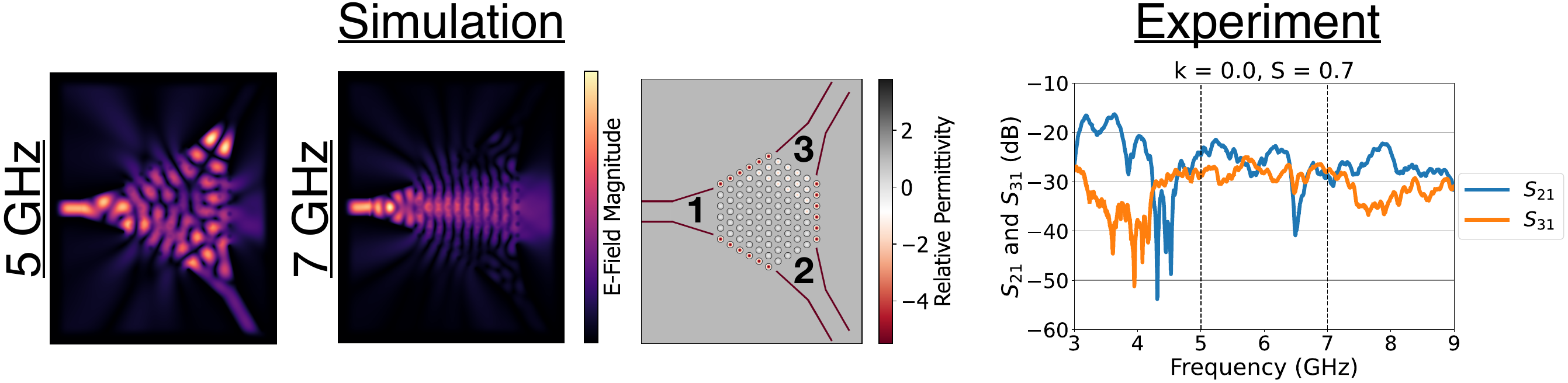}
		\caption{Field simulations $|\textbf{E}_z|$ and optimal device domains $\Re[\varepsilon]$ (left) along with experimental transmission spectra (right) for the PMM device with our best attempt at an \textit{a priori} human-designed 5 / 7 GHz demultiplexer. The vertical dashed black lines indicate the design frequencies, and the fit parameters $k$ and $S$ are chosen according to the best values for the devices conceived via inverse design.}
		\label{fig:ManDemult}
	\end{figure}

In figure \ref{fig:ManDemult}, we see the simulated and experimental results when a 5 / 7 GHz demultiplexer is designed by hand by creating what would function as low-index bridges to the correct output port for either design frequency. The general principle is that port 3 is protected by a region of sources that have an average plasma frequency $\sim 6$ GHz, so those elements serve as a reflective barrier to the 5 GHz source but are transparent to the 7 GHz source. The edges of the domain are set to their maximum plasma density to discourage leakage. $k=0.0$, and $S=0.7$ were chosen as the fitting parameters to limit the voltage of the edge elements an avoid damaging the discharges. In both the simulation and the experiment, the device doesn't appear to steer the 7 GHz source to either output, and there is only slight isolation between the ports in the correct direction at 5 GHz. The experimental spectrum is fairly unremarkable without obviously meaningful structures at any frequency.

\section{Discussion \& Conclusion}\label{sec:disc}

The results presented in the previous section show that although we can reliably get somewhat strong performance by optimizing our devices \textit{in-silico}, there are ultimately some shortcomings. In most cases, the configurations actually resulted in better performance at frequencies other than the design frequency as seen with the 6 GHz beam steering device in Fig. \ref{fig:Wvg} and the 5.5 / 6 GHz demultiplexer in Fig. \ref{fig:Demult}. These cases show that up to $\sim$40 dB isolation is achievable in practice at various frequencies throughout the frequency domain, indicating that the device has a high degree of dynamic range (possibly due to some manner of Fabry-P\'erot resonance effect), but the \textit{in-silico} optimizer was only able to produce 10-15 dB isolation at the design frequencies. 

While there can be a few reasons for these shortcomings, we attribute them to simplifications made in the simulations. The \textit{in-silico} optimization procedure uses a 2D domain with plasma elements of varying levels of complexity. The most sophisticated plasma element model that includes both a non-uniform plasma density profile and collisional damping failed to produce the best performance in all but one case, illustrating that we still lack the detailed knowledge of the discharge properties that are needed to produce robust results at design conditions. Furthermore, the system is not entirely electromagnetically isolated as assumed in the simulations. To avoid interferences due to scattering from the surroundings of the PMM, experiments are carried out in a custom-fabricated anechoic chamber, but even this isn't entirely sufficient. In addition, the finite size of the tubes brings the electrodes and the wiring into play, both of which undoubtedly interact with the source fields. Unfortunately, a 3D simulation that takes many of these non-ideal conditions into account is far too costly to use within an inverse design scheme.

We also found that the mapping between the power supply settings (voltage or current) and the plasma frequency of the discharge tubes, facilitated by the electron-energy distribution solver, BOLSIG+, tends to overestimate the plasma density, where only the lowest density case according to the BOLSIG+ calculations is compatible with the experimental results ($k=0$). One interesting departure from the other results regarding the experimental mapping is the 5 / 7 GHz demultiplexer that performs best for the collisional plasma element model at $k=1.0$, $S=1.0$, indicating that when using that element model, the optimal plasma frequencies from the \textit{in-silico} optimization are necessarily larger. Because the elements are lossy, the optimizer may be biased to higher plasma densities to encourage scattering off of the plasma elements instead of propagation through them. It's also possible that the collisionality is stronger in the experimental device at higher power (the collision frequency is assumed to be 1 GHz in the \textit{in-silico} collisional plasma model), therefore, to match the simulated result one has to decrease the plasma frequency in the experiment. Also, since the device does not function at precisely the intended frequencies, it is possible that this configuration was simply stumbled upon by chance. Much of these challenges can be avoided by \textit{in-situ} optimization.

The attempt at designing the device intuitively for objectives such as the demultiplexer is an illustration of how we cannot hope to outperform inverse design, particularly so when our model of the physical device is not precise. Despite the shortcomings discussed above, the devices still show strong evidence that the \textit{in-silico} optimization can result in performance that at least addresses the design criteria, if not at a high degree of efficiency. Designing by intuition is therefore not an option; but of course we have also shown here that the \textit{in-silico} inverse design procedure fails to fully utilize the dynamic range of the system elements. Luckily, since our device is composed entirely of plasma elements, we can perform the inverse design process entirely \textit{in-situ}.

Fully \textit{in-situ} inverse design addresses almost all the shortcomings described above and also opens up many exciting opportunities. We no longer have to be concerned with the fidelity of the modeling since we will be using the physical device to perform the optimization. Since we find that the device is capable of $\sim$50 dB isolation at various frequencies, we would expect to be able to achieve this and likely more since those cases were obtained accidentally. This also presents an opportunity to refine our element model. Once a high-performance design is achieved via \textit{in-situ} inverse design, the parameters can be fed into a simulation domain where we modify the plasma model until we see the same performance, allowing us to more thoroughly understand the physics of our plasma sources. Moving to fully \textit{in-situ} optimization also removes some of the limitations on our geometry. As mentioned before, 3D geometry would be far too computationally costly to use for \textit{in-silico} inverse design, but it incurs no extra cost in \textit{in-situ} inverse design. Thus, we could use device structures like that of ref. \cite{tunableppc}. Even for this 2D configuration, each iteration in the optimization procedure can take 10 minutes with a 64-core workstation, while the \textit{in-situ} iterations an be limited to mere seconds in practice regardless of the geometry. We have also shown in prior work that our plasma sources have a pronounced gyrotropic response when magnetized \cite{LucMag, HosMag}, but materials with anisotropic permittivity tensors like magnetized plasma can't be modeled with our simulation tool. By using a large Helmholtz configuration, entire PMM devices can be magnetized to take advantage of the very rich physics inherent to magnetized plasmas.

In conclusion, we show in this study that PMM devices can be optimized \textit{in-silico} to perform beam steering and demultiplexing at a level of performance that is not possible to obtain using conventional intuitive design methods. Ultimately, the \textit{in-silico} inverse design process has shortcomings that lead to a failure to utilize the full potential of the PMM configuration. Since the PMM device can be reconfigured in seconds or less, the inverse design process can be performed entirely \textit{in-situ}, side-stepping the computational cost and modeling inaccuracies associated with \textit{in-silico} inverse design, thereby promising better performance.

\begin{acknowledgments}
The authors would like to thank Dr. Benjamin Wang for his assistance in procuring the experimental components for this study. This research is supported by the Air Force Office of Scientific Research through a Multi-University Research Initiative (MURI), Grant No. FA9550-21-1-0244, with Dr. Mitat Birkan and Dr. Arje Nachman as the Program Managers. J.A.R. acknowledges support from the Charles H. Kruger Stanford Graduate Fellowship.
\end{acknowledgments}

\bibliography{refs}

\end{document}